# Diode and selective routing functionalities controlled by geometry in current-induced spin-orbit torque driven magnetic domain wall devices

*Elena M. Ștețco[1,2], Traian Petrișor Jr.[1], Ovidiu A. Pop[2], Mohamed Belmeguenai[3], Ioan M. Miron[4], Mihai S. Gabor[1,\*]*

[1] Centre for Superconductivity, Spintronics and Surface Science, Physics and Chemistry Department, Technical University of Cluj-Napoca, Str. Memorandumului, 400114 Cluj-Napoca, Romania

[2] Applied Electronics Department, Technical University of Cluj-Napoca, Str. Memorandumului, 400114 Cluj-Napoca, Romania

[3] Université Sorbonne Paris Nord, LSPM, CNRS, UPR 3407, F-93430 Villetaneuse, France

[4] Université Grenoble Alpes, CNRS, CEA, Grenoble INP, SPINTEC, Grenoble, France

ABSTRACT

Research on current-induced domain wall (DW) motion in heavy metal/ferromagnet structures is crucial for advancing memory, logic, and computing devices. Here, we demonstrate that adjusting the angle between the DW conduit and the current direction provides an additional degree of control over the current-induced DW motion. A DW conduit with a 45° section relative to the current direction enables asymmetrical DW behaviour: for one DW polarity, motion proceeds freely, while for the opposite polarity, motion is impeded or even blocked in the 45° zone, depending on the interfacial Dzyaloshinskii-Moriya interaction strength. This enables the device to function as a DW diode. Leveraging this velocity asymmetry, we designed a Y-shaped DW conduit with one input and two output branches at +45° and -45°, functioning as a DW selector. A DW injected into the junction exits through one branch, while a reverse polarity DW exits through the other, demonstrating selective DW routing.

INTRODUCTION

The electric manipulation of magnetic textures is a key area of research for the development of next-generation memory, logic, and in-memory computing devices [1-7]. In racetrack memory and logic devices, data is encoded within continuous strips comprising a series of magnetic domain walls (DWs), which are precisely displaced by in-plane electric current pulses [8,9]. A particularly significant focus is on perpendicularly magnetized heavy-metal (HM)/ferromagnetic (FM) metallic multilayers, which exhibit broken inversion symmetry and large spin-orbit coupling (SOC). Within these structures, an in-plane current injected into the HM layer generates both damping-like (DL) and field-like (FL) torques on the magnetization of the FM layer [10,11]. Moreover, the SOC induced Dzyaloshinskii-Moriya interaction (DMI) at the HM/FM interface [12] imposes a Néel DW configuration with in-plane DW magnetization aligned parallel to the current, thereby maximizing the efficiency of the damping-like torque in driving the DWs [13-15].

In straight strips, pulsed current-induced displacement of DWs with alternating polarities progresses in a synchronized, stepwise manner. Additionally, reversing the current direction results in an oppositely directed but symmetrical displacement of the DWs [14]. For both racetrack memories and logic devices, breaking the symmetry of alternating polarity DW displacement is of considerable interest, as it introduces additional degrees of control over the current-induced DW motion. Upon the application of relatively long current pulses the DW will tilt due to the canting of the DW magnetization, affecting the current induced domain wall dynamics [16]. This feature can be used in curved, bent or Y-shape junctions to create an asymmetry in the current induced motion of alternating polarities DWs [17-19].

In this study, we demonstrate that by separating the DW conduit from the current strip and adjusting the angle between the DW conduit and the current direction, it is possible to control current-induced DW motion for short pulses when no significant DW tilt is expected. By carefully tailoring the shape of the DW conduit and the interfacial DMI, we achieve the functionality of a DW diode: the DW moves in one direction under the influence of the current, while motion in the opposite direction is hindered or even blocked. Additionally, we show that Y-shaped DW conduits function as DW selective routers: DWs with one polarity follow one branch of the junction, while DWs with the opposite polarity follow the other branch.

RESULTS AND DISCUSSION

The current-induced chiral DW motion in HM/FM structures is a dynamic phenomenon, which is dependent on the angle between the direction of the current and of the DW motion [20]. To qualitatively understand this behavior and how it can be harnessed for the development of chiral DW devices, we employ a simplified model (elaborated in more details in the Supporting Information, section S1). Let us consider an up/down DW situated in the middle of a perpendicularly magnetized HM/FM strip. Initially, in the absence of any current and for strong enough DMI, the DW magnetization ($m$) aligns along the DMI effective field ($H_{DMI}$), pointing towards the negative $Ox$ axis, as indicated schematically in Fig.1(b)-i. When a charge current ($J_c$) is applied through the strip the DL torque $\tau_{DL} = -\gamma a_j m \times (m \times p)$ will act on the DW magnetization. The torque depends on the charge current density (through $a_j$, defined in the Supporting Information, section S1), on $m$, and on the spin-current polarization vector $p = e_J \times n$, which is transverse to both the direction of the charge current density ($e_J = J_c/J_c$) and the direction perpendicular to the strip $n = (0,0,1)$. The action of $\tau_{DL}$ is to rotate $m$ in-plane at an angle $\alpha$ away from the DMI axis (Fig.1(b)-ii). The associated DMI torque $\tau_{DMI} = -\gamma m_{DW} \times H_{DMI} = \gamma H_{DMI}(sin\,\alpha)k$ will rotate the magnetization out-of-plane, causing the DW to move. In principle, the DW velocity depends on the charge current density and the projection of the DW magnetization on the charge current direction, $v \propto J_C cos(\alpha - \beta)$ [13,21]. Therefore, by applying a current $J_C$ at various $\beta$ angles relative to the strip, it is possible to either minimize or maximize the DW velocity. This is shown in Fig.1(a), which depicts the velocity of an up/down DW calculated using micromagnetic simulations for an ideal strip (see Supporting Information, section S2). During simulations periodic boundary conditions were assumed along the Oy axis. This ensures that the DW has no tilt, and it remains always parallel with the *Oy* axis

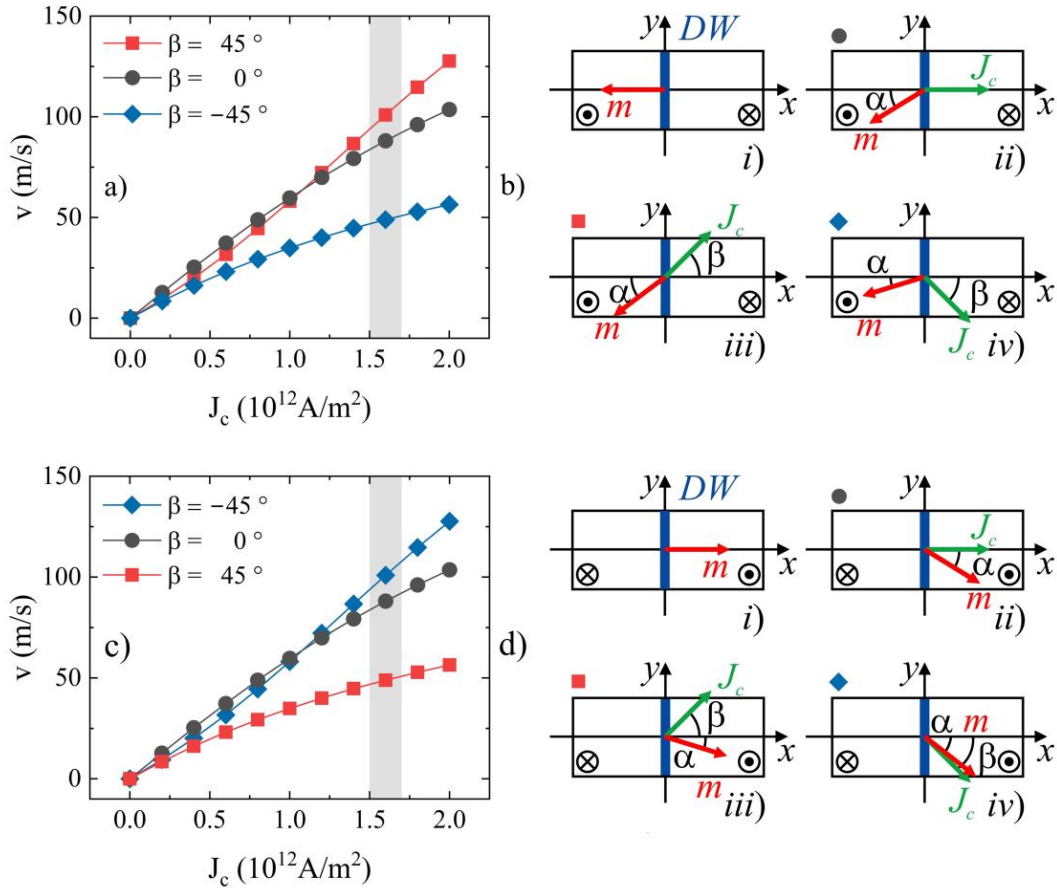

**Figure 1**(a) Up/down DW velocity derived from micromagnetic simulations for an ideal strip subjected to current pulses of 10 ns width and amplitudes up to $2 \times 10^{12}$ A/m², assuming a DMI constant of 1.1 mJ/m². The pulses are applied at $\beta = 45°$, $0°$ and $-45°$ with respect to the *Ox* axis. For pulses above $1 \times 10^{12}$ A/m², the DW velocity is maximized at $\beta = 45°$ and minimized at $\beta = -45°$. (b) Schematic representation of the DW magnetization $m$ relative to the charge current direction $J_c$ for the current pulses indicated by the shaded area in (a). In the absence of any current, $m$ aligns along the DMI axis. When current is applied, the DL torque tilts the DW magnetization counterclockwise. At $\beta = 45°$, $J_c$ and $m$ are relatively well aligned resulting in maximum DW velocity. At $\beta = 0°$ the alignment is reduced and the DW velocity decreases,

while at $\beta = -45°$, minimal alignment leads to minimum velocity. (c) Down/up DW velocity calculated under the same conditions as in (a). Contrary to (a), for pulses above $1 \times 10^{12}$ A/m², the DW velocity is maximized at $\beta = -45°$ and minimized at $\beta = 45°$. (d) Schematic representation of the DW magnetization $\boldsymbol{m}$ relative to the charge current direction $\boldsymbol{J_c}$ for the current pulses indicated by the shaded area in (c). In the absence of any current, $\boldsymbol{m}$ aligns along the DMI axis. When current is applied, the DL torque tilts the DW magnetization clockwise. At $\beta = -45°$, $\boldsymbol{J_C}$ and $\boldsymbol{m}$ are relatively well aligned resulting in maximum DW velocity. At $\beta = 0°$ the alignment is reduced and the DW velocity decreases, while at $\beta = 45°$, minimal alignment leads to minimum velocity."

and perpendicular to the $Ox$ axis. Current pulses of density up to $2 \times 10^{12}$ A/m² and 10 ns width are applied at 45°, 0° and −45° degrees with respect to the $Ox$ axis. A DMI constant $|D| = 1.1$ mJ/m² was used. At relatively low currents, because $\boldsymbol{\tau}_{DL}$ is small, the magnetization rotation angle $\alpha$ is close to zero. Consequently, as seen in Fig. 1(a), the highest velocity occurs when the current is applied along the $Ox$ axis ($\beta = 0$), where the current direction and the DW magnetization are most closely aligned compared to the other angles ($\beta = 45°$ and $\beta = -45°$). In contrast, at higher currents, the magnetization rotation angle $\alpha$ becomes significant. For instance, with current density pulses of $1.6 \times 10^{12}$ A/m² the DW magnetization rotation angles $\alpha$, shown in Fig. 1(b)ii-iv, indicate that at $\beta = 45°$, the current direction and the DW magnetization are most closely aligned, resulting in maximum velocity. However, when the current is applied along the $Ox$ axis ($\beta = 0$), there is less collinearity between the current direction and the DW magnetization, which results in lower velocity. Furthermore, at $\beta = -45°$, the current direction and the DW magnetization are least aligned, and the velocity is minimum.

At the same current density, due to the symmetry, the behavior reverses for a down/up domain: the velocity is maximized at $\beta = -45°$ and minimized at $\beta = 45°$, while at $\beta = 0$, the velocity matches that of the up/down DW (Fig. 1(c) and (d)).

Additional micromagnetic simulations (see Supporting Information, section S2) indicate that the geometry-induced velocity asymmetry is not restricted to $\beta = -45°$ and $\beta = 45°$, but is observable when the current is applied at any angle other than zero relative to the strip axis. Moreover, the simulations reveal that as the width of the DW conduit decreases, the velocity asymmetry not only persists but also increases as the width is reduced into the nanometer range. This indicates that the observed velocity asymmetry is robust and scalable across various DW conduit geometries.

The asymmetry in DW velocity when the current is applied at 45° and −45° relative to the strip suggests that, in practice, by selecting appropriate material parameters, it is feasible to maximize the DW velocity for one current direction while minimizing it, or even blocking DW motion entirely, for the opposite direction. This behavior is corroborated by micromagnetic simulations on realistic systems [22] (see Supporting Information, section S2).

To experimentally demonstrate the geometry-induced velocity asymmetry, we fabricated two types of thin film structures, as detailed in Supporting Information, section S3. One structure has relatively low DMI (LD), while the other has relatively high DMI (HD). The LD structure consists of Si/SiO$_2$/Ta(25)/Pt(60)/CFB(7)/Pd(2)/Pt(16), and the HD structure comprises Si/SiO$_2$/Ta(25)/Pt(60)/CFB(6)/Pd(6)/Pt(12). Here, the numbers in parentheses indicate the thickness of each layer in angstroms, with CFB representing the Co$_{60}$Fe$_{20}$B$_{20}$ alloy. The thicknesses of the CFB layers were adjusted in both structures to achieve PMA with roughly similar anisotropy fields (see Supporting Information, section S4). Here, an in-plane charge current is converted by the Pt layer into a perpendicular spin current and/or spin accumulation, which induces DW motion in the CFB layer via the spin-orbit torques (SOTs) [9,13-15,23]. The role of the top Pd/Pt bilayer is to modulate the interfacial DMI of the CFB layer. In these types of structures, the DMI is given mainly by the bottom Pt/CFB interface, while the DMI of the top CFB/(Pd)/Pt interface will add destructively. By varying the thickness of the Pd layer the iDMI induced by the top Pt layer can be gradually screened [24] enabling precise control over the total DMI. Brillouin light scattering (BLS) experiments (see Supporting Information, section S5) revealed an effective DMI constant of 0.48 mJ/m$^2$ for the LD structure, and 1.1 mJ/m$^2$ for HD structures.

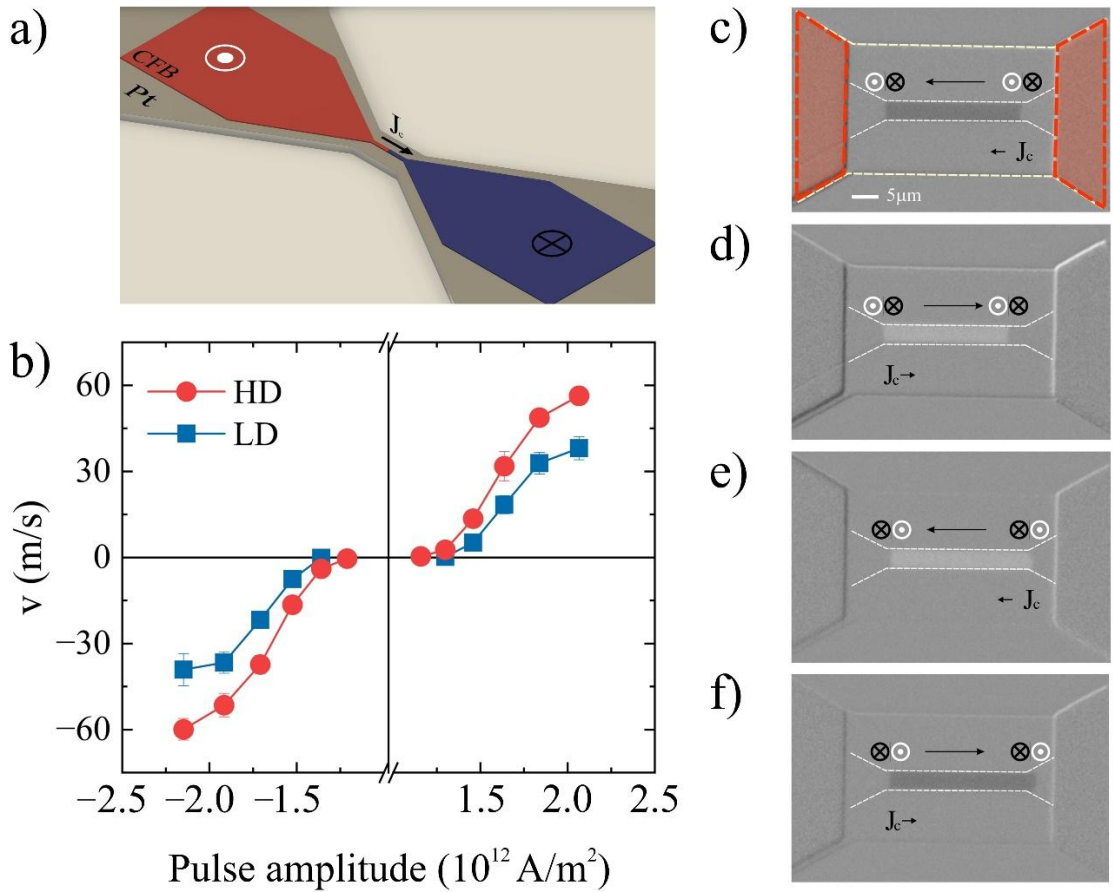

"**Figure 2** (a) Schematic representation of a straight CFB DW conduit patterned on top of the Pt current strip. Red (blue) regions denote up (down) magnetized domains. An up/down DW is positioned in the middle of the conduit. (b) DW velocity as a function of pulse amplitude for HD and LD samples. Beyond a certain threshold, the velocity undergoes a rapid increase, characteristic of thermally activated DW motion, marking the transition to the flow DW motion regime. The onset of DW velocity saturation is also observed, occurring for higher values for the HD samples compared to the LD ones, consistent with the SOT-DMI current-induced DW motion mechanism. Points are average values of at least 10 measurements on four patterned devices, using two sets of each sample deposited in different runs. Error bars correspond to the standard deviation. (c)-(f) Differential MOKE images showing DW propagation after the application of current pulses. Dark contrast indicates up/down (down/up) DW motion to the left (right), while bright contrast indicates up/down (down/up) DW motion to the right (left). DWs move in the direction of the applied current, consistent with left-handed DW chirality. In (c), the DW conduit boundaries are highlighted by white dashed lines, the Pt current strip by yellow dashed lines, and the red shaded area marks the electrical contacts."

To control the charge current direction relative to the DW, the samples were patterned as 3 µm wide DW conduits on top of 26 µm wide Pt strips using photolithography and Ar ion milling (see Supporting Information, section S3). We first analyze the current induced DW motion in a straight DW conduit, shown schematically in Fig. 2(a). The DWs were nucleated and placed in their initial position by an appropriate series of out-of-plane magnetic field pulses. Voltage pulses of varying amplitude and a width of 2.75 ns were applied on the Pt strip, and the resulting DW displacement was observed using wide-field magneto-optical Kerr effect (MOKE) microscopy. A voltage pulse of 25 V is equivalent to a current density of $1.2 \div 1.3 \times$ $10^{12}$ Am$^{-2}$ assuming that the current is flowing only through the current strip. This assumption overestimates the current through the Pt layer, providing only an upper bound. Although a more accurate estimation would require employing more sophisticated methods [25], determining the exact current density is not critical for the objectives of our study. Figures 2(c)-(f) display typical differential MOKE images obtained by subtracting the initial image from the one recorded after the application of the current pulses. The dark contrast indicates the motion of an up/down (down/up) DW towards the left (right) side, while the light contrast shows the motion of an up/down (down/up) DW towards the right (left) side. The DWs move in the direction the applied current, as expected for the left-handed DW chirality [13]. Figure 2(b) shows the DW velocity as a function of the pulse amplitude. Both types of samples show similar behavior. Above a certain threshold, the velocity increases rapidly indicating the transition from the creep to the flow DW motion regime [26]. The onset of DW velocity saturation is also observed, at a higher value for the HD sample compared to the LD one, as expected for the SOT-DMI current induced DW motion mechanism [13,15,27].

In the second step, the DW conduit was designed with a section forming a $\pm 45°$ angle relative to the current direction (Fig. 3(a)). This geometry allows the observation of DW motion within the same sample when the current is either parallel to or at a $\pm 45°$ angle relative to the DW conduit. Figure 3(c) shows MOKE images indicating the displacement of a down/up DW under the action of positive and negative pulses in the LD sample. Initially, a down/up DW is nucleated in the lower left part of the structure, and a series of $N = 280$ positive pulses with an amplitude of $2.1 \times 10^{12}$ Am$^{-2}$ are applied (Fig. 3(c), top). The DW moves under the action of the current through the entire conduit up to the right top side of the structure. Subsequently, the same number of pulses with amplitude $2.0 \times 10^{12}$ Am$^{-2}$ are applied in negative direction. In this case, the DW motion is blocked in the 45° zone of the DW conduit. This occurs immediately after the DW passes the corner of the conduit, when the DW angle approaches 45° relative to the current strip (Fig. 3(c), middle; see also Supporting Information Video SV1). To confirm that the DW is indeed blocked, a background MOKE image was recorded and extracted from the image taken after application of additional $10 \times N$ pulses and shown in the bottom of part Fig. 3(c). No discernible DW displacement is noticeable. For geometric reasons, a down/up DW in the 45° zone under positive pulses corresponds to the situation described in Fig. 1(d-iv), while under negative pulses it corresponds to the one in Fig. 1(b-iv). A similar situation occurs for an up/down DW (Fig. 3(d)). The DW moves through the entire structure from the top right side to the bottom left side with the application of $N = 280$ negative pulses (Fig. 3(d), top). When positive pulses are applied, the DW is blocked in the 45° zone (Fig. 3(d), middle; see also Supporting Information Video SV2). Further application of positive pulses does not result in a discernable DW displacement (Fig. 3(d), bottom). Similarly, a up/down DW in the 45° zone under positive pulses corresponds to the situation described in Fig. 1(b-iv), while under negative pulses it corresponds to the one in Fig. 1(d-iv).

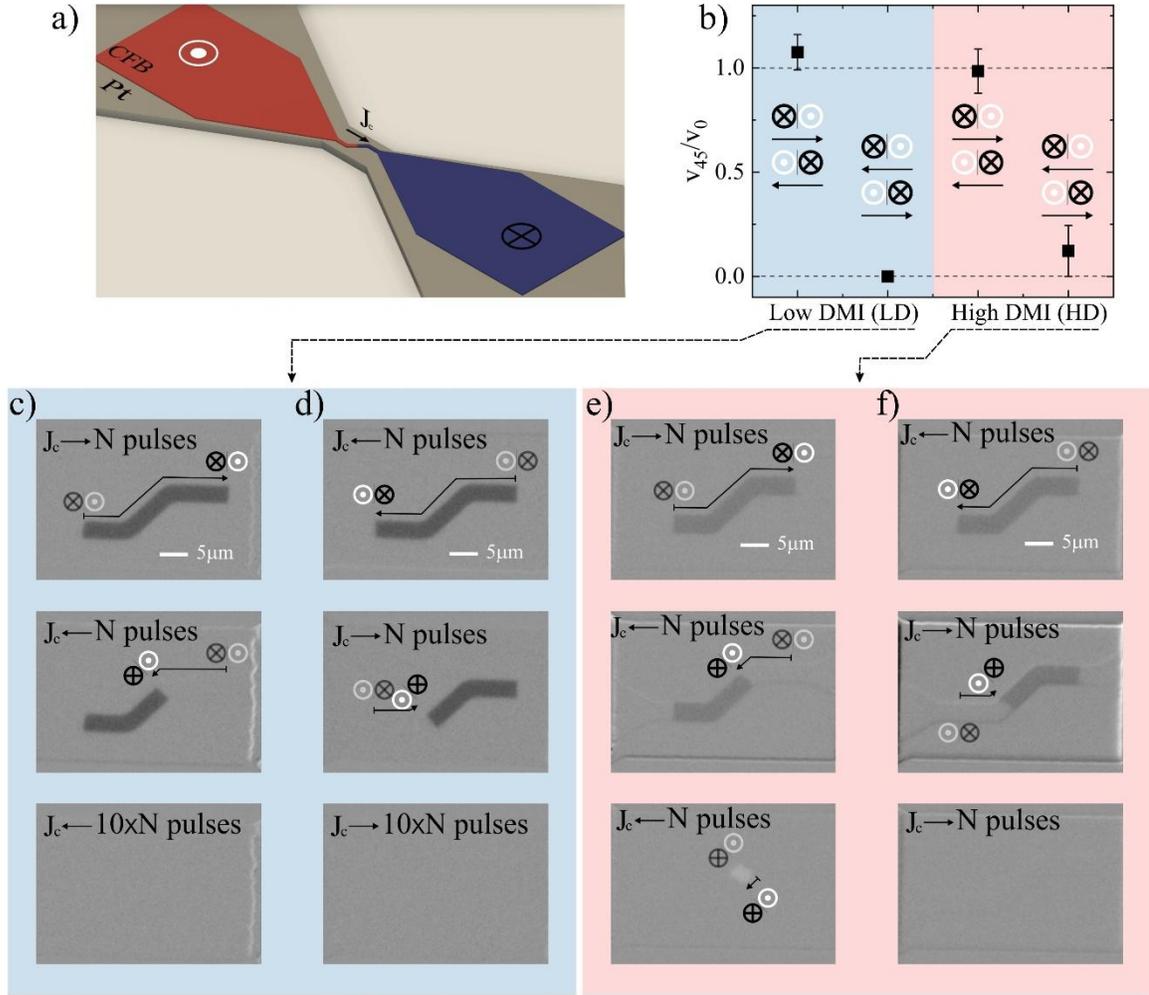

**Figure 3** (a) Schematic representation of a CFB DW conduit patterned on the Pt current strip, with a section tailored to form a $\pm 45°$ angle relative to the current direction. The red region indicates an up-magnetized domain, while blue region indicates a down-magnetized domain, with an up/down DW located in the middle of the 45° zone. (b) DW velocity in the 45° zone normalized to the velocity in the straight zone for the LD and HD samples. The points represent average values, with error bars extending to the most distant outliers. LD sample: (c) A down/up DW moving under the action of $N = 280$ positive (top part) and negative pulses (middle part). For positive pulses the DW moves through the entire structure, and for negative pulses the motion is blocked in the 45° zone, even after further application of $10 \times N$ pulses (bottom part).

(d) An up/down DW moving under the action of $N = 280$ negative (top part) and positive pulses (middle part), with a similar behavior as in (c). HD sample: (e) A down/up DW moving under the action of $N = 175$ positive (top part) and negative pulses (middle part). For negative pulses the DW motion is not blocked in the 45° zone, but its velocity is diminished (bottom part). (f) An up/down DW moving under the action of $N = 175$ negative (top part) and positive pulses (middle part). For positive pulses, the DW motion is blocked in the 45° zone, and further application of $N$ positive pulses does not set the DW in motion (bottom part)."

For the HD sample, the DW behavior differs. A down/up domain moves through the entire structure from the lower left to the upper right with the application of $N = 175$ positive pulses (Fig. 3(e), top). When the same number of negative pulses are applied, the DW reaches the 45° zone (Fig. 3(e), middle; see also Supporting Information Video S3), but it is not blocked; its velocity is only diminished. Further application of $N$ negative pulses result in a discernible DW displacement in the 45° zone (Fig. 3(e), bottom; see also Supporting Information Video SV4). As in the case of the LD sample, an up/down domain nucleated in the top right side of the conduit will move through the entire structure when $N = 175$ negative pulses are applied (Fig. 3(f), top). The DW motion is reversed for positive pulses and blocked in the 45° zone (Fig. 3(f), middle part; see also Supporting Information Video SV5). Further application of negative pulses does not result in additional DW displacement (Fig. 3(f), bottom; see also Supporting Information Video SV6).

Figure 3(b) shows the DW velocity in the 45° zone normalized to the velocity in the straight zone for the LD and HD samples. The points represent values averaged for up to four different patterned devices and for two series of each sample deposited in different runs, with error bars extended to include the furthest outliers. For the LD sample, the velocity in the 45° zone for a down/up DW moving to the right or an up/down DW moving to the left is slightly higher than the velocity in the straight zone. Conversely, a down/up DW moving to the left or an up/down DW moving to the right is always blocked in the 45° zone. For the HD samples, the velocity in the 45° zone for a down/up DW moving to the right or an up/down DW moving to the left is roughly the same as the velocity in the straight zone. However, the velocity of a down/up DW moving to the left or an up/down DW moving to the right is, on average, approximately one order of magnitude lower than the velocity in the straight zone. The fact that the normalized velocity in the 45° zone for a down/up DW moving to the right or an up/down DW moving to the left is slightly higher for the LD sample compared to the HD one aligns with our simple model and micromagnetic simulations, where a larger velocity asymmetry is expected for the lower DMI case for the same current density (see sections S1 and S2 of the Supporting Information). For a down/up DW moving to the left or an up/down DW moving to the right the driving-force behind the DW motion decreases, and the DW velocity is largely determined by the pinning potential landscape and thermal activation [13-15]. In the case of LD lower DMI sample the driving-force decrease is large enough so no DW displacement can be detected even after application of $10 \times N$ pulses. For the higher DMI HD sample, the DW moves fast between pinning sites, but experiences extended waiting periods at each site, resulting in an overall one order of magnitude decrease in velocity (see also Supporting Information S5). The observation that one type of DW can move unobstructed in one direction while being hindered and even blocked in the reverse direction demonstrates the realization of DW diode functionality in the device.

Due to the specific geometry of our device, the out-of-plane Oersted field generated by the current passing through the Pt strip affects DW motion differently in the two straight sections. If it inhibits DW motion in the first straight section, it promotes motion in the second, and vice versa, potentially influencing the overall functionality of the device. However, as demonstrated in Supporting Information, section S7, the effect of the out-of-plane Oersted field is negligible. Even if this were not the case, it would not significantly affect the functionality of the device, as the out-of-plane Oersted field goes to zero at the center of the 45° zone and would not affect the DW velocity in that critical region.

An important question is whether the observed asymmetric behavior is intrinsic, or an extrinsic, material-dependent feature related to DW pinning. To address this, we deposited analogous samples using Co instead of CFB. Co has a higher depinning field, and its crystalline structure results in more defects compared to the amorphous CFB. Consequently, Co is expected to exhibit a higher density of pinning sites with stronger pinning potentials relative to CFB. As detailed in Supporting Information, section S7, the Co samples show a comportment analogous to that of the CFB samples. This supports the idea that observed asymmetric behavior is intrinsic rather than being a material-dependent feature.

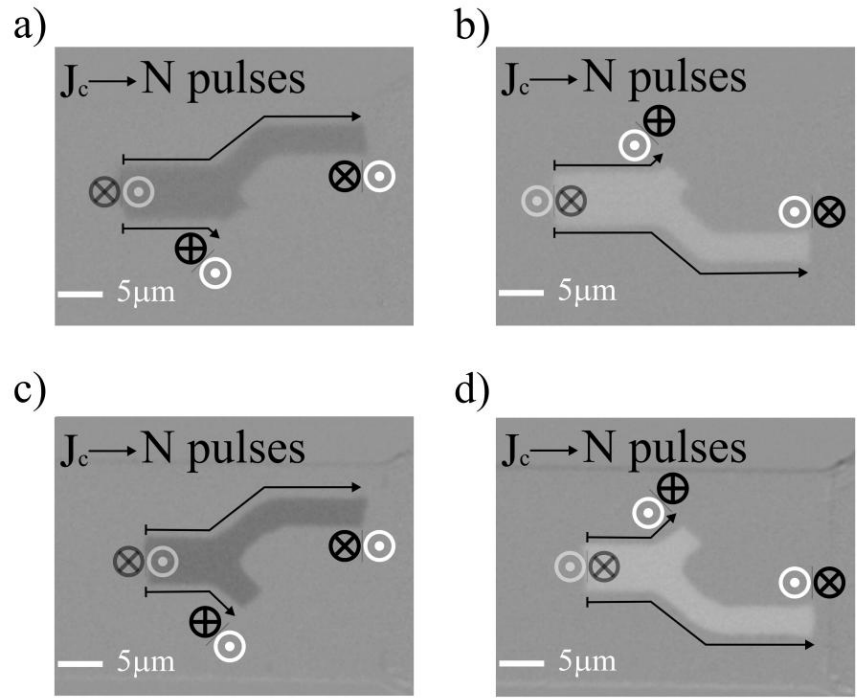

"**Figure 4** A Y-shaped DW selector with two branches oriented at −45° and 45° relative to the current direction. LD structure: (a) A down/up DW moving under the action of $N = 275$ positive pulses. The DW moves freely in the upper branch, while in the lower branch the DW motion is blocked. (b) An up/down DW showing mirror symmetry relative to (a), the DW is blocked in the upper branch and moves freely in the lower branch. HD structure: The down/up (c) or up/down (d) DWs show similar behavior with the application of $N = 150$ pulses, as in the in case of the LD structure. The only difference from the LD case is that the down/up (up/down) DW is not completely blocked in the lower (upper) branch, but it moves with reduced velocity."

The DW velocity asymmetry between up/down and down/up DWs in the 45° zone can be exploited to construct DW selectors. To achieve this, we designed devices where the straight DW conduit splits into two branches oriented at −45° and 45° relative to the current direction. For the LD structure, a down/up domain is nucleated in the left part of the straight conduit, followed by the application of a series of $N = 275$ pulses with an amplitude of $2.1 \times 10^{12}$ Am$^{-2}$ (Fig. 4(a)). As the DW reaches the bifurcation, it enters both branches. The DW motion is obstructed in the lower branch, where the DW conduit and current form a 45° angle. In contrast, the DW moves freely in the upper branch, where the DW conduit and current form a $-45°$ angle, allowing it to reach the second straight zone. It is interesting to observe that the up/down DW behavior exhibits mirror symmetry (Fig. 4(b)). In this case, the DW is blocked in the upper branch, where the DW conduit and current form a $-45°$ angle, and moves freely in the lower branch, where the DW conduit and current form a 45° angle. The HD structure shows a similar behavior with the application of $N = 150$ pulses (Fig. 4(c) and (d)). The primary distinction is that the down/up (up/down) DW is not entirely blocked in the lower (upper) branch but progresses at a significantly reduced velocity (see also Supporting Information Video SV7). Consequently, our device demonstrates DW selective routing: down/up DWs traverse the upper branch freely, while up/down DWs traverse the lower branch freely.

Micromagnetic simulations (Supporting Information, section S2), supported by the experimental data, indicate that the velocity asymmetry depends on the DMI strength. Additionally, simulations reveal that adjusting the angles of the two output branches in the Y-shaped DW conduit can modulate this asymmetry. Specifically, the asymmetry increases as the angle between the branches increases, providing an additional means to control the DW velocity asymmetry. Consequently, the Y-shaped device can function as a DW splitter with asymmetric and tunable outputs, enabling precise control over DW trajectories in complex network architectures, analogous to in-plane magnetized DW networks [28]. This enhanced control over DW propagation could have significant implications for a diverse array of DW-based applications, including the development of unconventional computing schemes [29].

CONCLUSIONS

In summary, we have shown that adjusting the angle between the DW conduit and the current strip provides additional geometric control over current-induced DW motion. Micromagnetic simulations highlighted the role of DL torque in creating an asymmetry in DW velocity when the current is applied at angles of 45° and −45° relative to the DW conduit. This velocity asymmetry was leveraged by designing the DW conduit with a 45° region. MOKE microscopy experiments demonstrated that both low-DMI (LD) and high-DMI (HD) structures exhibit DW diode-like behavior. For LD structures, a down/up (or up/down) DW moves freely through the entire structure under a series of positive (or negative) pulses but reversing the current blocks the DW motion in the 45° zone. In HD samples, the behavior is similar, except the DW motion in the 45° zone is not completely blocked but with a rather significantly reduced velocity. Finally, we utilized the DW velocity asymmetry to construct DW selectors. In the Y-shaped DW conduit with one input and two, 45° and −45°, output branches, a DW injected into the junction will exit through one branch, while a DW of reverse polarity will exit through the other branch, thereby demonstrating selective routing of DWs.

METHODS

**Device fabrication**

The LD structure consisting of Si/SiO$_2$/Ta(25)/Pt(60)/CFB(7)/Pd(2)/Pt(16), and the HD structure comprising Si/SiO$_2$/Ta(25)/Pt(60)/CFB(6)/Pd(6)/Pt(12), where the numbers in parentheses indicate the thickness of each layer in angstroms, and with CFB representing the Co$_{60}$Fe$_{20}$B$_{20}$ alloy, were deposited on thermally oxidized Si substrates using d.c. magnetron sputtering at an argon pressure of 1.5 mtorrr. For current-induced DW motion experiments, the samples were patterned as 3 μm wide DW conduits on top of 26 μm wide Pt strips using UV photolithography and Ar-ion milling. The main steps of the device patterning are given in the Supporting Information.

**Micromagnetic simulations**

For a better understanding of current-induced DW motion mechanisms we use micromagnetic simulations performed with MuMax3 code. The system is discretized in 2 nm x 2 nm x 0.6 nm cells. The following magnetic parameters are used: saturation magnetization $Ms = 1.37 \times 10^6$ A/m, exchange coupling constant $A = 10 \times 10^{-12}$ J/m, perpendicular uniaxial anisotropy constant $K_u = 1.525 \times 10^6$ J/m$^3$, Gilbert damping parameter $\alpha_G = 0.6$, the spin-Hall angle $\theta_{SHE} = 0.07$ and the ratio between the field-like and the damping like SOTs $\xi = 0.2$. The DMI was varied from $|D| = 0.5$ mJ/m$^2$ up to $|D| = 1.1$ mJ/m$^2$. Both ideal and disordered systems are considered.

**MOKE microscopy measurements**

The MOKE images were recorded using a custom-built wide-field polar Kerr effect microscope. DW motion is driven by current pulses generated by a Picosecond Pulse Labs 2600C pulse generator, which supplies high amplitude pulses up to 45 V with a fast rise time of less than 250 ps. The initial magnetic state of the device is prepared by application of appropriate out-of-plane magnetic field pulses. Then, voltage pulses of varying amplitude and a width of 2.75 ns were applied within the Pt stripe and the DW motion is observed using the MOKE microscope.

ASSOCIATED CONTENT

**Supporting Information.**

Model of the DW magnetization dynamic tilting, micromagnetic simulations, details on the samples growth and microfabrication, magnetic properties and DMI characterization, tracking the DW motion in the 45° zone (PDF).

(Movie SV1) MOKE movie demonstrating the diode functionality of the LD device. A down/up DW is displaced by the application of 280 positive and negative pulses (MP4).

(Movie SV2) MOKE movie demonstrating the diode functionality of the LD device. An up/down DW is displaced by the application of 280 negative and positive pulses (MP4).

(Movie SV3) MOKE movie demonstrating the diode functionality of the HD device. A down/up DW is displaced by the application of 175 positive and negative pulses (MP4).

(Movie SV4) MOKE movie showing the displacement of a down/up DW within the HD device by the application of 750 negative pulses. The DW is not blocked in the 45° zone, its velocity is reduced by approximately one order of magnitude (MP4).

(Movie SV5) MOKE movie demonstrating the diode functionality of the HD device. An up/down DW is displaced by the application of 175 negative and positive pulses (MP4).

(Movie SV6) MOKE movie demonstrating the diode functionality of the HD device. An up/down DW is displaced by the application of 175 negative and positive pulses. The DW is blocked in the 45° zone (MP4).

(Movie SV7) MOKE movie demonstrating the DW selector functionality of the HD device. A down/up and an up/down DW is displaced by the application of 150 positive pulses. The down/up DW follows the upper branch, while the up/down DW follows the lower branch of the Y-shaped device (MP4).

AUTHOR INFORMATION

**Corresponding Author**

* Mihai S. Gabor - Centre for Superconductivity, Spintronics and Surface Science, Physics and Chemistry Department, Technical University of Cluj-Napoca, Str. Memorandumului, 400114 Cluj-Napoca, Romania. Email: mihai.gabor@phys.utcluj.ro

**Author Contributions**

M.S.G coordinated the research. E.M.S., T.P.Jr., M.S.G grew the films and patterned the devices. E.M.S., T.P.Jr,. O.A.P., I.M.M. and M.S.G. conceived the experiments. E.M.S. performed the experiments. M.B. measured the DMI by BLS. M.S.G supervised and wrote the manuscript with contributions from E.M.S. All the authors discussed the results and participated in preparing the manuscript. All authors have given approval to the final version of the manuscript.

**Funding Sources**

This work was supported by a grant of the Romanian Ministry of Education and Research, CNCS - UEFISCDI, project number PN-III-P4-ID-PCE-2020-1853, within PNCDI III. T.P.Jr. acknowledges funding from a grant of the Romanian Ministry of Education and Research, CNCS-UEFISCDI, project number PN-III-P1-1.1-TE-2021-1777, within PNCDI III. I.M.M. acknowledges funding for this work from the European Research Council (ERC) under the European Union's Horizon 2020 research and innovation programs: ERC-StG Smart Design (638653) and ERC-PoC SOFT (963928).

**Notes**

The authors declare no competing financial interest

# Supporting Information

**Diode and selective routing functionalities controlled by geometry in current-induced spin-orbit torque driven magnetic domain wall devices**

*Elena M. Ştețco[1,2], Traian Petrişor Jr.[1], Ovidiu A. Pop[2], Mohamed Belmeguenai[3], Ioan M. Miron[4], Mihai S. Gabor[1,*]*

[1] Centre for Superconductivity, Spintronics and Surface Science, Physics and Chemistry Department, Technical University of Cluj-Napoca, Str. Memorandumului, 400114 Cluj-Napoca, Romania

[2] Applied Electronics Department, Technical University of Cluj-Napoca, Str. Memorandumului, 400114 Cluj-Napoca, Romania

[3] Université Sorbonne Paris Nord, LSPM, CNRS, UPR 3407, F-93430 Villetaneuse, France

[4] Université Grenoble Alpes, CNRS, CEA, Grenoble INP, SPINTEC, Grenoble, France

**S1. Chiral domain wall magnetization dynamic tilting in the spin-orbit torques - Dzyaloshinskii–Moriya interaction model**

**S2. Micromagnetic simulations**

**S3. Growth and device fabrication**

**S4. Magnetic properties of the LD and the HD stacks**

**S5. Details on the interfacial Dzyaloshinskii–Moriya interaction measurements.**

**S6. Tracking the motion of the DW in the 45° zone**

**S7. Oersted field effect**

**S8. Different ferromagnetic material**

## S1. Chiral domain wall (DW) magnetization (m) dynamic tilting in the spin-orbit torques (SOT) - Dzyaloshinskii–Moriya interaction (DMI) model

Let us consider an up/down domain wall (DW) aligned along the $Oy$ axis in a heavy metal/ferromagnetic (HM/FM) strip, which exhibits perpendicular magnetic anisotropy (PMA), as schematically illustrated in Fig. S1. The charge current ($J_C$) is applied at an angle $\beta$ relative to the $Ox$ axis.

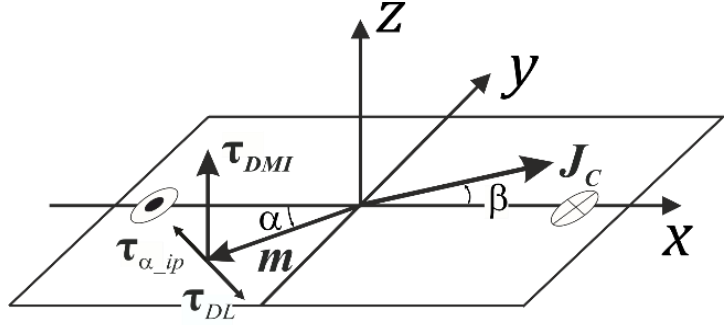

**Figure S1**. Schematic representation of the different torques acting on the DW magnetization $m$.

The magnetization dynamics is governed by the Landau-Lifshitz-Gilbert (LLG) equation incorporating the spin-orbit torques (SOTs) [1-3]:

$$\frac{\partial m}{\partial t} = -\gamma m \times H_{eff} + \alpha_G m \times \frac{\partial m}{\partial t} + \tau_{DL} + \tau_{FL}, \tag{S1}$$

where $m$ is the normalized magnetization, $\gamma$ the gyromagnetic ratio and $\alpha_G$ the damping constant. The effective field $H_{eff} = -\frac{1}{\mu_0 M_s}\frac{\delta E}{\delta m}$ is derived from the energy density of the system and includes contributions from the exchange, dipolar, anisotropy and Dzyaloshinskii–Moriya interaction (DMI). The third and fourth terms of the LLG equation are the damping-like and field-like torques. The damping-like torque is given by $\tau_{DL} = -\gamma a_j m \times (m \times p)$ and the field-like torque by $\tau_{FL} = -\gamma b_j m \times p$, where $p = \text{sign}(\theta_{SHE}) e_J \times n$ is the spin current polarization unit vector, which is transverse to both the direction of the charge current density ($e_J = J_C/J_C$) and the direction perpendicular to the strip $n = (0,0,1)$. The $\mu_0 a_j = \left|\frac{\hbar \theta_{SHE} J_C}{2 e M_s t_{FM}}\right|$ term is the damping-like effective field strength, where $\mu_0$ is the vacuum magnetic permeability, $\hbar$ is the reduced Planck constant, $e$ is the elementary charge, $t_{FM}$ is the thickness of the FM layer. For a given set of material parameters, the damping-like effective field strength is proportional to the current density $J_c$. The field-like effective field strength is given by $\mu_0 b_j = \xi \mu_0 a_j$ [4]. In the above relations $\theta_{SHE}$ is the spin-Hall angle and $\xi$ is the ratio between the field-like and the damping like SOTs. Due to the damping-like torque symmetry, the associated damping-like effective field $\boldsymbol{H}_{DL} = a_j \boldsymbol{m} \times \boldsymbol{p}$ is always locally perpendicular to the magnetization direction. The field-like torque effective field $\boldsymbol{H}_{FL} = b_j \boldsymbol{p}$ is equivalent to a magnetic field transverse to the direction of the charge current.

The normalized DW magnetization writes as

$$\boldsymbol{m} = -m_x \boldsymbol{i} - m_y \boldsymbol{j}, \tag{S2}$$

where:

$$\begin{cases} m_x = \cos \alpha, \\ m_y = \sin \alpha, \end{cases}$$

are the components along $Ox$ and $Oy$, with the $\alpha$ angle defined in Fig. S1.

The charge current density $\boldsymbol{J}_c$ is defined as

$$\boldsymbol{J}_c = (J_x \boldsymbol{i} + J_y \boldsymbol{j}) J_C, \tag{S3}$$

where:

$$\begin{cases} J_x = \cos \beta, \\ J_y = \sin \beta, \end{cases}$$

are the charge current normalized components along the $Ox$ and $Oy$ axis and the $\beta$ angle is defined in Fig. S1.

Assuming a current applied at an angle β relative to the $Ox$ axis direction and a positive spin-Hall angle $\theta_{SHE}$, the spin current polarization unit vector is given by

$$\boldsymbol{p} = \begin{vmatrix} \boldsymbol{i} & \boldsymbol{j} & \boldsymbol{k} \\ J_x & J_y & 0 \\ 0 & 0 & 1 \end{vmatrix} = J_y \boldsymbol{i} - J_x \boldsymbol{j} = \sin \beta \, \boldsymbol{i} - \cos \beta \, \boldsymbol{j}. \tag{S4}$$

In this geometry, the damping-like effective field is oriented out-of-plane and writes as

$$\boldsymbol{H}_{DL} = a_j \boldsymbol{m} \times \boldsymbol{p} = a_j \cos(\beta - \alpha) \boldsymbol{k}. \tag{S5}$$

The corresponding torque acts on the $\boldsymbol{m}$ by rotating it in-plane with an angle $\alpha$ relative to the $-Ox$ axis direction and it is given by

$$\boldsymbol{\tau}_{DL} = -\gamma \boldsymbol{m} \times \boldsymbol{H}_{DL} = \gamma a_j \cos(\beta - \alpha)[\sin \alpha \, \boldsymbol{i} - \cos \alpha \, \boldsymbol{j}]. \tag{S6}$$

This essentially indicates that the damping-like torque depends on the charge current density, through $a_j$, the charge current direction, through the angle $\beta$, and on the orientation of the DW magnetization, through the angle $\alpha$.

The field-like effective field writes as

$$\boldsymbol{H}_{FL} = b_j \boldsymbol{p} = b_j \sin\beta \, \boldsymbol{i} - b_j \cos\beta \, \boldsymbol{j}. \tag{S7}$$

The corresponding field-like torque is given by

$$\boldsymbol{\tau}_{FL} = -\gamma \boldsymbol{m} \times \boldsymbol{H}_{FL} = -\gamma b_j \cos(\beta - \alpha) \boldsymbol{k}. \tag{S8}$$

Similarly to the damping-like torque, the field-like torque depends on the charge current density, through $b_j$, charge current direction, through the angle $\beta$, and on the orientation of the DW magnetization, through the angle $\alpha$.

Since the DMI stabilize a left-handed chirality of DW, the effective field associated with the DMI, $\boldsymbol{H}_{DMI}$, is oriented along $-Ox$ axis direction, $\boldsymbol{H}_{DMI} = H_{DMI}(-1,0,0)$ and its strength is given by

$$H_{DMI} = \frac{D}{\mu_0 M_s \Delta}, \tag{S9}$$

where $D$ is the absolute DMI constant and $\Delta = \sqrt{A/K_{eff}}$ represents the domain wall width, determined by the exchange stiffness $A$ and the effective anisotropy $K_{eff} = K_u - \frac{\mu_0 M_s^2}{2}$ [5].

The associated torque, $\boldsymbol{\tau}_{DMI}$, writes as

$$\boldsymbol{\tau}_{DMI} = -\gamma \boldsymbol{m} \times \boldsymbol{H}_{DMI} = \gamma H_{DMI}(\sin\alpha) \boldsymbol{k}, \tag{S10}$$

which depends on the DMI strength and the orientation of the DW magnetization, through the angle $\alpha$.

The combined effect of $\boldsymbol{\tau}_{FL}$ and $\boldsymbol{\tau}_{DMI}$ is to rotate the magnetization out-of-plane and to impose the DW velocity.

The dissipative torque associated with the DW motion is given by [6]:

$$\boldsymbol{\tau}_\alpha = -\gamma \boldsymbol{m} \times \boldsymbol{H}_\alpha, \tag{S11}$$

where

$$\boldsymbol{H}_\alpha = -\frac{\alpha_G}{\gamma} \frac{\partial \boldsymbol{m}}{\partial t}. \tag{S12}$$

The domain wall moves as a result of the action of $\boldsymbol{\tau}_{DL}$, $\boldsymbol{\tau}_{FL}$ and $\boldsymbol{\tau}_{DMI}$, thus by combining the above equations one can write:

$$\boldsymbol{\tau}_\alpha = \alpha_G \boldsymbol{m} \times \frac{\partial \boldsymbol{m}}{\partial t} = \alpha_G \boldsymbol{m} \times (\boldsymbol{\tau}_{FL} + \boldsymbol{\tau}_{DL} + \boldsymbol{\tau}_{DMI}), \tag{S13}$$

which gives the in-plane component for the damping torque

$$\boldsymbol{\tau}_{\alpha_{ip}} = \alpha_G \gamma [H_{DMI} \sin\alpha - b_j \cos(\alpha - \beta)][-\sin\alpha\, \boldsymbol{i} + \cos\alpha\, \boldsymbol{j}]. \tag{S14}$$

The in-plane component of the damping torque opposes the damping-like torque, and when $\tau_{DL} = \tau_{\alpha\_ip}$, i.e.

$$\cos(\alpha - \beta) = \alpha_G \left[ \frac{H_{DMI}}{a_j} \sin\alpha - \xi \cos(\alpha - \beta) \right], \tag{S15}$$

they cancel each other and impose a certain $\alpha$ angle. For a given set of material parameters, $\alpha$ angle depends on the charge current density $J_c$ and angle $\beta$ at which the current is applied. If the field-like torque is much smaller than the damping-like torque ($\xi \approx 0$), the above relation becomes:

$$\cos(\alpha - \beta) = \frac{\alpha_G H_{DMI}}{a_j} \sin\alpha \tag{S16}$$

Using equation (S16) one can have a qualitative description for the steady-state $\alpha$ angle of the DW magnetization, during DW motion. In fig. S2 we plot the left-side and the right-side of the equation (S16), as a function of the $\alpha$ angle of the DW magnetization, for $J_c = 2.0 \times 10^{12}$ A/m$^2$, and a DMI value of $|D| = 1.1$ mJ/m$^2$. The other parameters are the identical to the ones used for micromagnetic simulations (see section S2). The continuous-line curves correspond to the left-hand side of the equation (S16) and the dotted-line curves correspond to the right-hand side term. For different current directions, the domain wall magnetization is rotated to different $\alpha$ angles. In case of $\beta = -45°$, the $\alpha$ angle is smallest, for $\beta = 0°$, the $\alpha$ angle is larger and for $\beta = +45°$, the $\alpha$ angle is the largest. The $\alpha$ angle values dependence on the current direction is in qualitative agreement with the micromagnetic simulations (Fig.1 from the main text), despite our over simplified model which does not consider all the aspects included in the full micromagnetic simulations. As indicated by equation (S16), the magnetization rotation $\alpha$ angle depends on the charge current. For lower currents the overall $\alpha$ angles are smaller. For instance, in the case of a charge current of $J_c = 0.6 \times 10^{12}$ A/m$^2$, the $\alpha$ angle is still smallest for $\beta = -45°$, but it is largest for $\beta = 0°$, and for $\beta = +45°$ it is intermediate. Another consequence of equation (S16) is that the rotation angle depends on the product $\alpha_G H_{DMI}$. These are material parameters which can be tuned in a certain range by a proper choice of the sample stack.

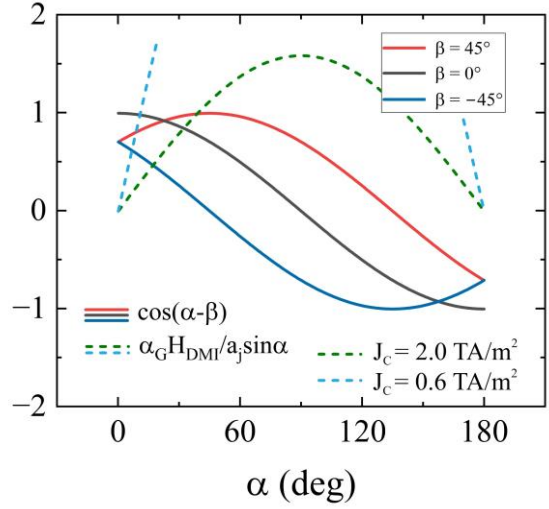

**Figure S2**. Plot of the left-hand side and right-hand side of equation S16.

With analogy with the field induced DW motion, in the steady state regime and for $H_{DL} \ll H_{DMI}$ the domain wall velocity is given by [5,6]:

$$v_{DW} = \frac{\gamma \Delta}{\alpha_G} |H_{DL}| \propto J_{SOT} \cos(\alpha - \beta). \quad (S17)$$

Is to be mentioned that a similar relationship was obtained in the case of full micromagnetic simulations [7,8]. The above equation implies that the DW velocity depends on the charge current density and the projection of the DW magnetization on the charge current direction. Consequently, the velocity is largest when the DW magnetization is aligned with the current and it decreases towards zero when they are transverse (Fig.1 from the main text).

**S2. Micromagnetic simulations**

The micromagnetic simulations were performed using MuMax³ [9-11]. The FM strip is discretized in 2 nm × 2 nm × 0.6 nm cells on top of a HM layer. We use the following typical parameters for a Pt/CFB bilayer matching our experimental results: saturation magnetization $Ms = 1.37 \times 10^6$ A/m, exchange coupling constant $A = 10 \times 10^{-12}$ J/m, perpendicular uniaxial anisotropy constant $K_u = 1.525 \times 10^6$ J/m³, Gilbert damping parameter $\alpha_G = 0.6$, the spin-Hall angle $\theta_{SHE} = 0.07$ and the ratio between the field-like and the damping-like SOTs, $\xi = 0.2$ [12].

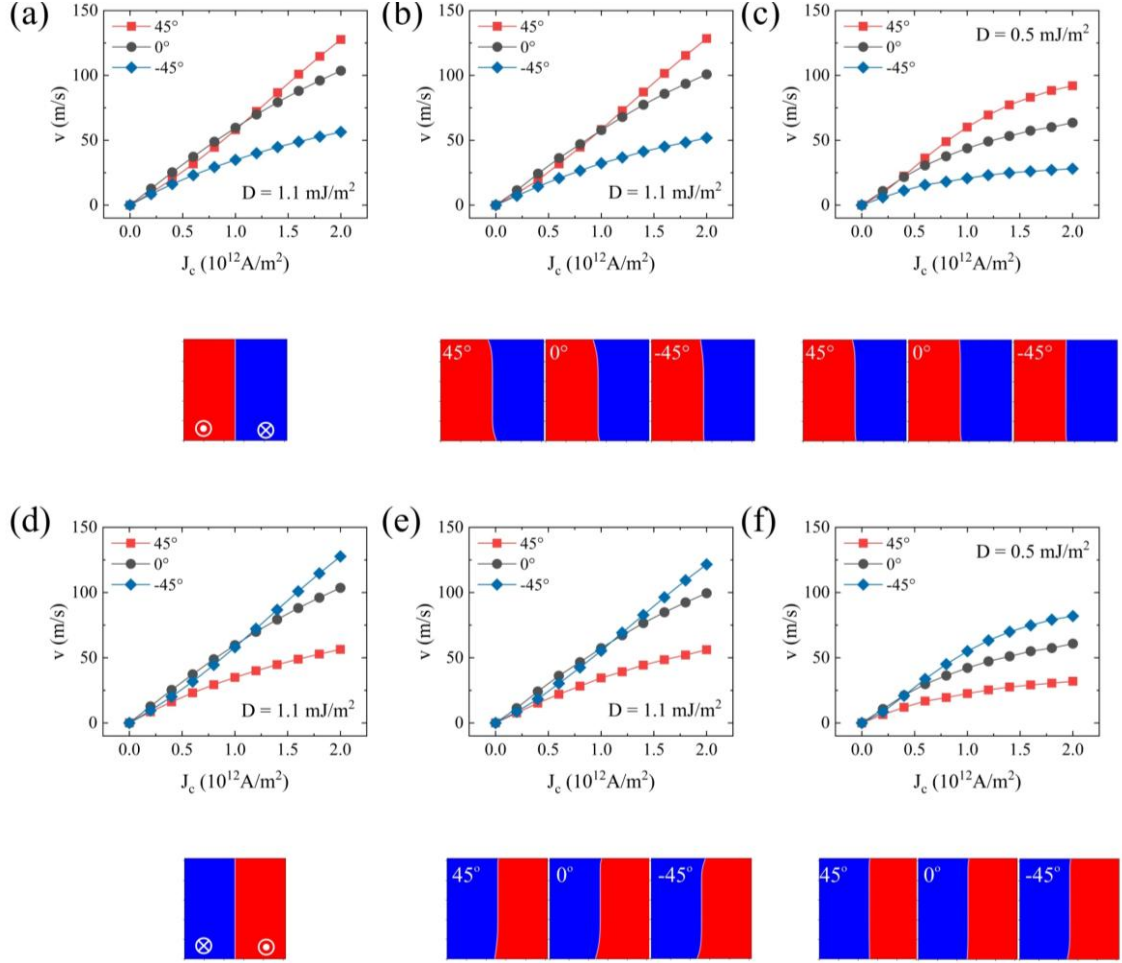

**Figure S3**. SOT driven DW velocity in presence of DMI calculated for an ideal strip subjected to 3 ns current pulses at angles $\beta = +45°$, $\beta = 0$ and $\beta = -45°$ from the strip axis. Three scenarios are presented: (a) an up/down DW and infinitely wide strip with $|D| = 1.1$ mJ/m², (b) an up/down DW in a 1.024 µm wide strip and a DMI of $|D| = 1.1$ mJ/m², and (c) an up/down DW in a strip of the same width and a DMI of $|D| = 0.5$ mJ/m². The lower part displays the corresponding DW profiles after the application of current pulses of $2.0 \times 10^{12}$ A/m². (d)-(f) show the same calculations as (a)-(c), except in the case of a down/up DW.

Figure S3 illustrates the velocity of an up/down DW calculated for an ideal strip subjected to 3 ns current pulses at angles $\beta = +45°$, $\beta = 0$ and $\beta = -45°$ from the strip axis. These calculations are shown for three different scenarios: an infinitely wide strip and a DMI of $|D| =$ 1.1 mJ/m² (Fig. S3(a)), a strip with a width of 1.024 µm and a DMI of $|D| = 1.1$ mJ/m² (Fig. S3(b)), and a strip of the same width and a DMI of $|D| = 0.5$ mJ/m² (Fig. S3(c)). In the lower part of the figures the corresponding DW profiles are shown following the application of current pulses of $2.0 \times 10^{12}$ A/m². For the infinitely wide strip (Fig. S3(a)), the DW remains straight and perpendicular to the strip, as expected, due to the assumption of periodic boundary conditions along the $Oy$ axis, simulating an infinite width system. In the absence of periodic boundary conditions, the DW undergoes slight deformation; the largest being observable for and $|D| = 1.1$ mJ/m² and the current applied at an angle $\beta = +45°$. No important tilt of the DW is observed, most likely due to the relatively large strip width and low current pulse duration [13]. The DW wall velocity shows roughly the same behavior in the case of 1.024 µm and infinite wide strips. The main distinction is observed between $|D| = 1.1$ mJ/m² and $|D| = 0.5$ mJ/m²; in the latter case, the velocity begins to saturate at lower currents, and the relative variation of the velocity for $\beta = +45°$ relative to $\beta = 0°$ is larger. This is expected considering that the in-plane DW magnetization rotation depends on the DMI field (section S1). The behavior of a down/up domain is similar. This is indicated in Fig. S3(d)-(f), where the same calculations are performed for a down/up DW. However, in this case the velocity is maximized at $\beta = -45°$ and minimized at $\beta = 45°$, while at $\beta = 0$, the velocity matches that of the up/down DW.

An important point to consider is the width of the strip for which the velocity asymmetry effect remains observable. We expect that the width of the DW conduit plays a role, as the tilting of the DW depends on the width of the strip [13]. Figure S4 illustrates the calculated velocity of an up/down DW for ideal strips with widths of 512 nm, 256 nm, and 128 nm, subjected to 3 ns current pulses at angles $\beta = +45°$, $\beta = 0$ and $\beta = -45°$ from the strip axis.

For a DMI strength of $|D| = 1.1$ mJ/m², as the width of the strip decreases, the tilting of the DW becomes more pronounced, particularly for $\beta = +45°$ and $\beta = 0°$. However, the velocity asymmetry persists and even increases as the width decreases, reaching its maximum value for the narrower simulated strip of 128 nm. A similar result is observed for a DMI of $|D| = 0.5$ mJ/m², although the DW tilting is less pronounced, as expected since DW tilting also depends on the strength of the DMI [13].

Our simulation results suggest that even as the width of the ferromagnetic strip decreases, the velocity asymmetry persists, indicating that the experimentally observed features should still be present even when the strip width is reduced to the nanometer range.

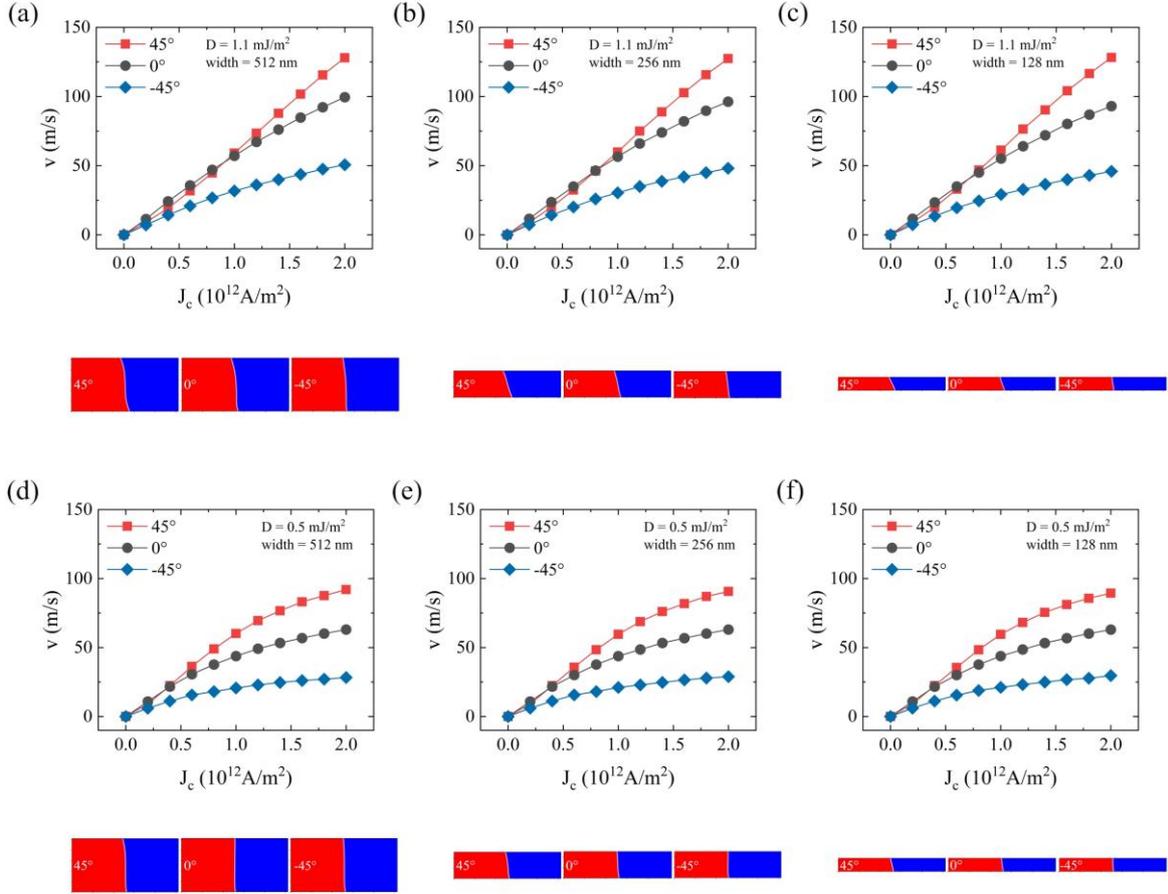

**Figure S4**. SOT driven velocity of an up/down DW calculated for strips of different widths, subjected to 3 ns current pulses, at angles $\beta = +45°$, $\beta = 0$ and $\beta = -45°$ from the strip axis, for (a)-(c) a DMI of $|D| = 1.1$ mJ/m², and (d)-(f) a DMI of $|D| = 0.5$ mJ/m². The lower part displays the corresponding DW profiles after the application of current pulses of $2.0 \times 10^{12}$ A/m².

As suggested by our simplified model in Section S1, the geometry-induced velocity asymmetry should be observable when the current is applied at any angle other than zero relative to the strip axis. Figure S5 (a) and (b) show the calculated velocities of an up/down DW for angles ranging from $\beta = +60°$ to $\beta = -60°$ from the strip axis. The calculations are presented for a strip with a width of 1.024 μm, with DMI strengths of |D|=1.1 mJ/m² and |D|=0.5 mJ/m², respectively. Figures S5 (c) and (d) display the corresponding velocities asymmetries, defined as $(v_\beta - v_0)/v_0$. The velocity asymmetry is DMI and current-dependent, and increases with β. This effect is more pronounced in the lower DMI structure, where the velocity begins to saturate at current densities above $10^{12}$ A/m². Beyond this current threshold, the velocity asymmetry becomes evident even at small angles and continues to grow as the angle increases.

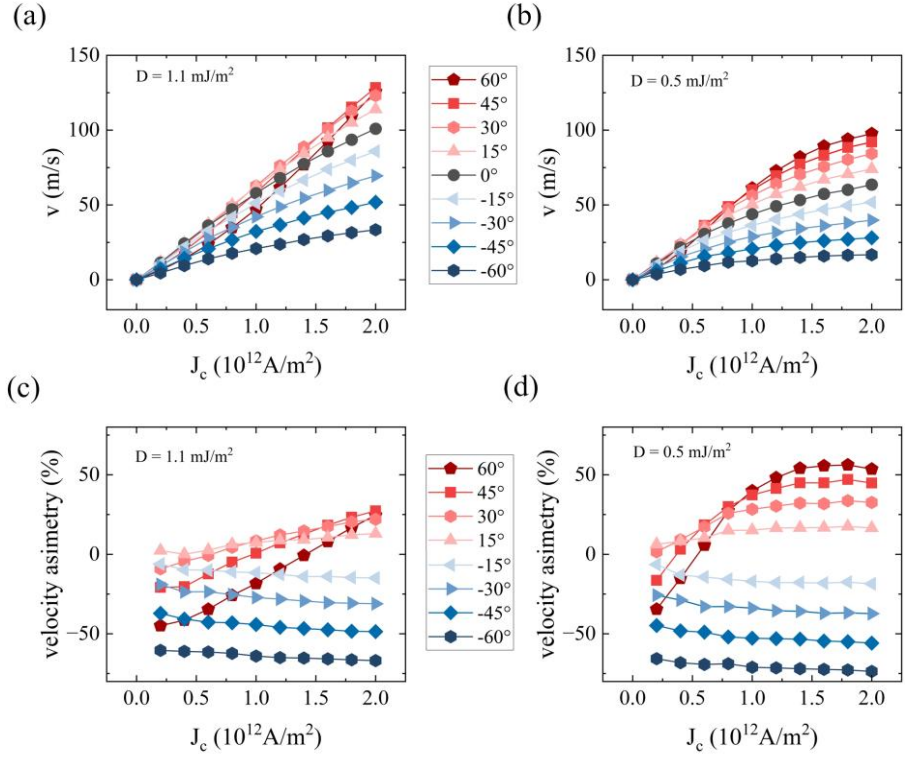

**Figure S5**. (a) and (b) SOT driven velocity of an up/down DW for angles ranging from $\beta = +60°$ to $\beta = -60°$ from the strip axis; (c) and (d) the corresponding velocity asymmetries, defined as $(v_\beta - v_0)/v_0$.

To simulate more realistic systems, we introduce disorder as local variations in the perpendicular anisotropy constant, within randomly generated grains structures with an average size of 5 nm. The magnitude of uniaxial magnetic anisotropy varies by up to 5% from one grain to another. Additionally, a small in-plane component of the uniaxial magnetic anisotropy is assumed, with its direction randomly generated and its magnitude not exceeding 5% of the out-of-plane component. To have statistical relevance, 40 randomly generated grain structures are simulated. A series of five consecutives 3 ns current pulses are applied. Between pulses, the energy of the system is minimized using the gradient descent method. Figure S6 shows the calculated velocity as a function of the applied pulse current density along with the DW profile following the application of $2.0 \times 10^{12}$ A/m² pulses. As expected, for a disordered system, there is a threshold current above which the DW starts to move. For larger pulse amplitudes, the main features found for the ideal system are present in the case of the more realistic structure: the velocity begins to saturate at lower currents, and the relative variation of the velocity for $\beta = +45°$ and $\beta = -45$ relative to $\beta = -0°$ is larger in the case of the lower DMI structure compared to the higher DMI one.

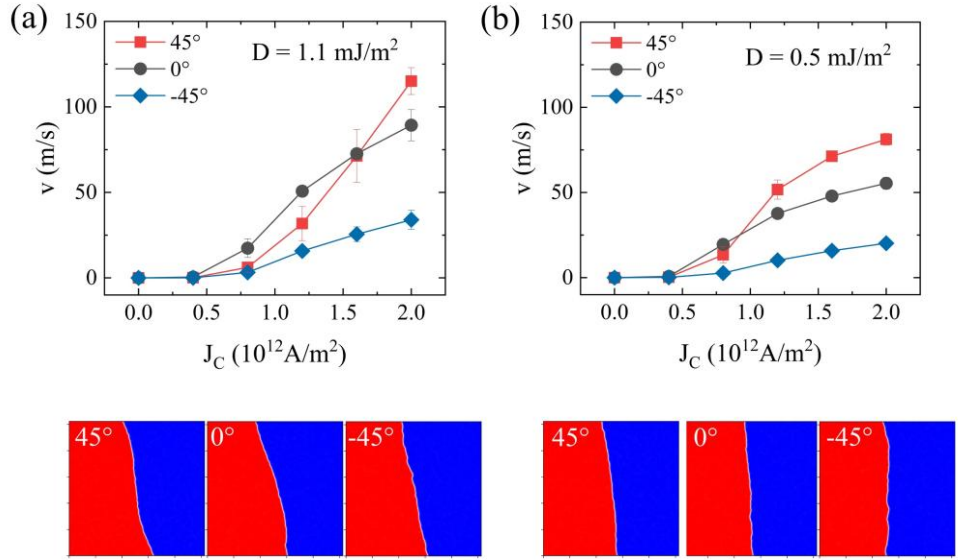

**Figure S6**. Calculated DW velocity as a function of applied pulse current density, together with typical DW profiles, after the application of $2.0 \times 10^{12}$ A/m² current pulses in the case of the granular system.

## S3. Growth and device fabrication

The low-DMI (LD) structure consisting of Si/SiO$_2$/Ta(25)/Pt(60)/CFB(7)/Pd(2)/Pt(16), and the high-DMI (HD) structure comprising Si/SiO$_2$/Ta(25)/Pt(60)/CFB(6)/Pd(6)/Pt(12), where the numbers in parentheses indicate the thickness of each layer in angstroms, and with CFB representing the Co$_{60}$Fe$_{20}$B$_{20}$ alloy, were deposited on thermally oxidized Si substrates using d.c. magnetron sputtering at an argon pressure of 1.5 mtorrr. The base pressure of the system was around $2 \times 10^{-8}$ Torr and the deposition rates around 0.1 nm/s.

For pulsed current experiments, the samples were patterned as 3 µm wide DW conduits on top of 26 µm wide Pt strips using UV photolithography and Ar-ion milling. The main steps of the device patterning are shown in Fig. S7. In a first lithography step the current line is defined (Fig. S7(a)). A dual photoresist process (LOR3A/Shipley 1813) was used to create an undercut in the photoresist mask, improving the edge roughness of the strip and limiting redepositions during the Ar-ion milling process. In the second lithography step, the DW conduit is defined on top of the current line (Fig. S7(b)) using a negative photolithography mask and the dual photoresist process, followed by the deposition of a 1.5 nm Ti hard mask. Next, Ar-ion milling is used to produce the DW conduit. The Ar-ion milling process was carried out in very small increments, complemented by MOKE microscopy observations, until the magnetic signal is lost outside the Ti hard mask protected area. Finally, using a negative photoresist (mA-N 1410) and lift-off technique, Ta(5 nm)/Cu(50 nm)/Pt(5 nm) electrical contacts are produced.

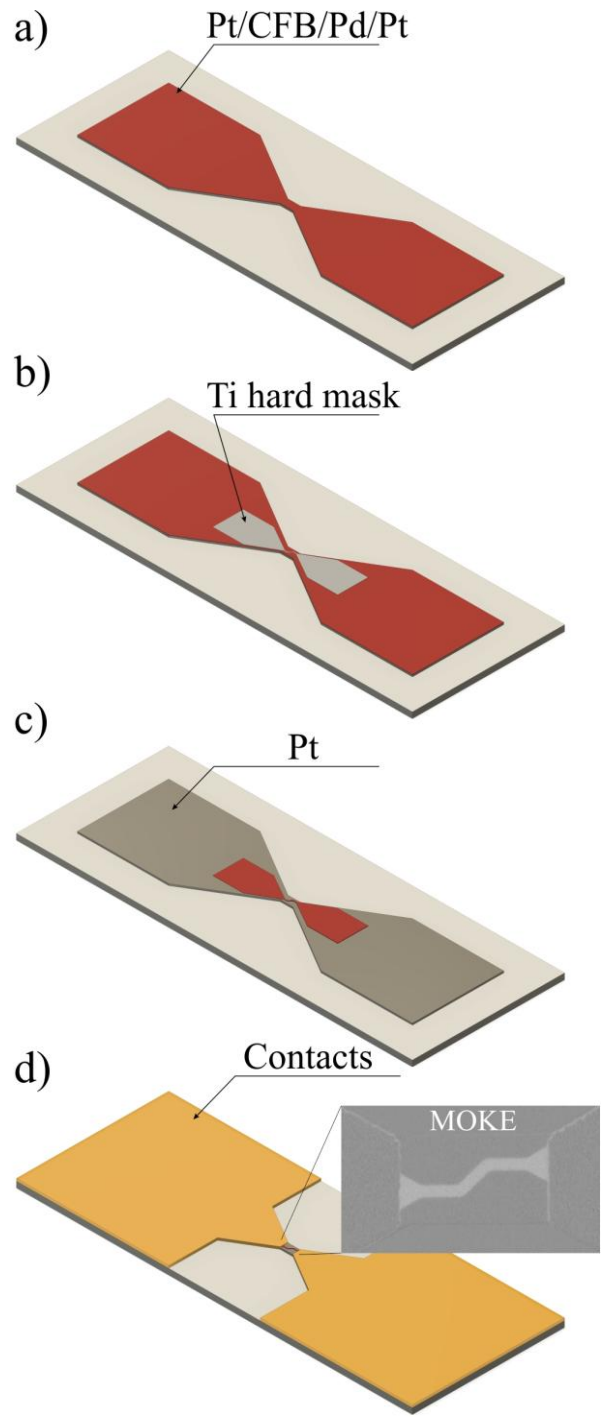

**Figure S7**. Schematic representation of the main device patterning steps. (a) Initial Ar-ion milling of the Pt/CFB/Pd/Pt stack, (b) deposition of the Ti 1.5 nm hard mask to define the magnetic conduit, (c) Ar-ion milling to produce the DW conduit, (c) deposition of the Ta/Cu/Pt electrical contacts. The inset shows a MOKE microscopy image of the DW conduit patterned on top of the Pt current stripe.

## S4. Magnetic properties of the LD and the HD stacks

Figure S8 presents hysteresis loops measured with the magnetic field applied in-plane (a) and perpendicular to the plane (b) for the LD - of Si/SiO$_2$/Ta(25)/Pt(60)/ CFB(7)/Pd(2)/Pt(16), and HD - Si/SiO$_2$/Ta(25)/Pt(60)/CFB(6)/Pd(6)/Pt(12) sample stacks. Both samples exhibit roughly the same saturation magnetization and demonstrate perpendicular magnetic anisotropy with comparable coercive and anisotropy fields. Notably, the LD structure exhibits a somewhat higher anisotropy field, due to the lower thickness of the CFB layer.

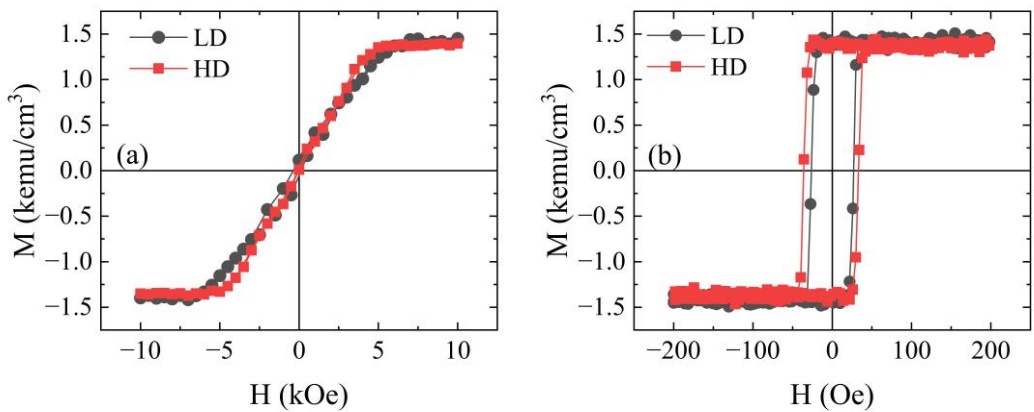

**Figure S8**. (a) In-plane and (b) out-of-plane hysteresis loops measured for the LD and HD samples,

## S5. Details on the interfacial Dzyaloshinskii–Moriya interaction measurements.

Brillouin light scattering (BLS) was employed to investigate the interfacial Dzyaloshinskii–Moriya interaction (iDMI) for the LD and HD stacks. Brillouin light scattering spectra were acquired, in the Damon-Eshbach configuration, for variable spin wave-vector (8.079 μm$^{-1}$≤$k_{sw}$ ≤20.45 μm$^{-1}$) and under a 10 kOe in-plane magnetic field, strong enough to saturate the samples in-plane. The spectra were fitted with Lorentzian functions to determine the Stokes (S) and anti-Stokes (aS) frequencies (see Fig. S9(a)), and to investigate the frequency difference ($\Delta F = F_S - F_{aS}$) versus $k_{sw}$. The frequency mismatch varies linearly with $k_{sw}$ for both samples, as shown in Fig. S9(b). The linear dependence of $\Delta F$ versus $k_{sw}$ is a signature of the iDMI contribution, especially for the ultrathin CFB films investigated here, and thus can be fitted by the relation $\Delta F = D_{eff} \frac{4\gamma}{2\pi M_s} k_{sw} = D_s \frac{4\gamma}{2\pi M_s t} k_{sw}$, where $D_{eff} = D_s/t$ with $D_s$ the iDMI surface constant, $M_s$ the magnetization at saturation and $\gamma/2\pi$=30.13 GHz/T the gyromagnetic ratio for CFB, measured previously [14]. This allowed us to obtain $D_{eff}$= -1.1 mJ/m² and -0.48 mJ/m² for the HD and LD systems, respectively.

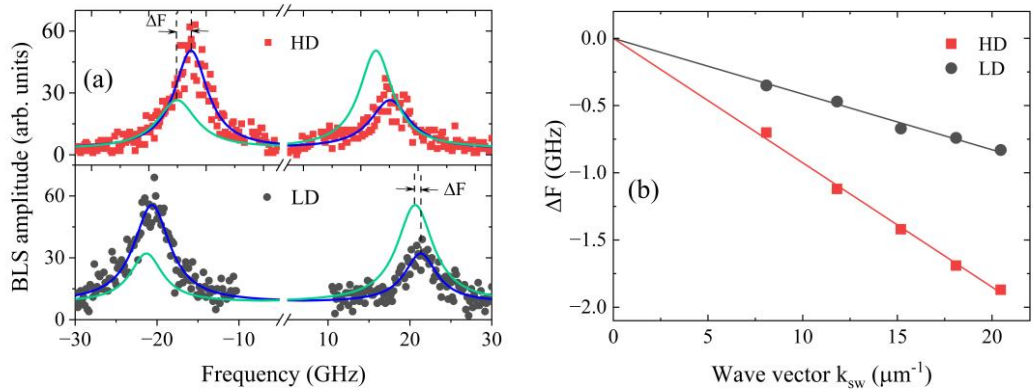

**Figure S9.** (a) BLS spectra measured for the HD and LD system at 10 kOe in-plane applied magnetic field and at a spin wave vector $k_{sw}$ = 20.45 μm$^{-1}$. Symbols refer to experimental data and solid lines are the Lorentzian fits. Fits corresponding to negative applied fields (green lines) are presented for clarity and direct comparison of the Stokes and anti-Stokes frequencies. (b) Variation of the frequency mismatch $\Delta F$ versus $k_{sw}$. Symbols are experimental data and solid lines are linear fits.

## S6. Tracking the motion of the DW in the 45° zone

To better understand the motion of the DW in the 45° zone for the HD samples, a video was recorded during the application of 750 negative pulses with an amplitude of $1.9 \times 10^{12}$ Am$^{-2}$ (supplementary video SV4). A delay of 100 ms was set between pulses. A custom image recognition software was used to track the DW motion. Figure S10 shows the DW displacement versus the number of recorded frames. The blue shaded areas correspond to the DW moving in the straight zones, while the white areas correspond to the DW moving in the 45° zone. In the straight zone, the DW moves with a constant velocity. Upon entering the 45° zone, the velocity drops to roughly one-third of its value in the straight zone, which aligns well with the micromagnetic simulations (S2). The DW maintains a relatively constant velocity until the first major pinning event occurs, indicated by the first arrow. The DW remains pinned for a relatively long period until a depinning event, indicated by the second arrow. The DW then continues in a pinning/depinning manner until a second major pinning event, indicated by the third arrow. After the last major depinning event, indicated by the fourth arrow, the DW enters the second straight zone and resumes moving at a constant velocity.

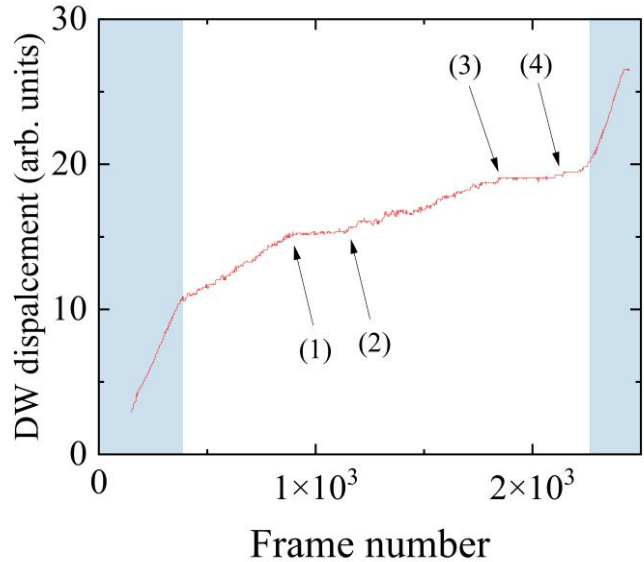

**Figure S10.** DW displacement versus the frame number for the HD system subjected to 750 negative voltage pulses.

**S7. Oersted field effect**

Due to the specific geometry of our devices, the out-of-plane Oersted field generated by the current passing through the Pt strip affects the DW motion differently based on its position. The impact varies depending on whether the DW is in the first straight section, the 45° zone, or the second straight section. For example, Fig. S11(a)(i) illustrates the motion of a down/up DW under positive current pulses. In the first straight section, located in the lower half of the Pt strip, the perpendicular Oersted field points into the strip, which promotes the DW motion in this region. Going through the 45° zone, the out-of-plane Oersted decreases and changes sign in the middle. In the second straight zone the perpendicular Oersted field points out of the strip and inhibits the DW motion in this region. By analyzing other cases shown in Fig. S11(a)(ii-iv), we conclude that the effect of the out-of-plane Oersted field is opposite in the two straight sections. If it inhibits DW motion in the first straight section, it promotes motion in the second, and vice versa. Therefore, the observable effect of the out-of-plane Oersted field is expected to be a change in DW velocity between the first and second straight zones. To compare the DW velocities in the two straight sections, we tracked the DW motion as described in section S6. Figures S11(c) and (d) show the DW displacement versus the number of recorded frames for both the HD and LD samples. For the LD sample, 280 current pulses were applied, while 175 pulses were applied for the HD samples. The blue-shaded areas represent the DW moving through the straight sections, while the white areas indicate the DW moving through the 45° zone. The DW displacements in the two straight zones were linearly fitted and the average slope ($\alpha_{med} = (\alpha_{szI} + \alpha_{szII})/2$) was extracted from the DW displacements values ($d_{DW} - \alpha_{med} \times$ frame number) and plotted in the lower part of the figures. It should be noted that the slopes of the linear fits correspond to DW velocity. As shown in the lower part of Fig. S11(c), for the HD samples, the dependence is roughly linear and horizontal, indicating that the DW velocity remains constant throughout the structure. In the case of the LD sample (Fig. S11(d)), the dependence is also linear and horizontal for the two straight zones, indicating a similar DW velocity in both. However, in the 45° zone, the dependence is linear and increasing, suggesting a higher DW velocity compared to the straight zones. This observation aligns with the data in Fig. 3(b) and the micromagnetic simulations.

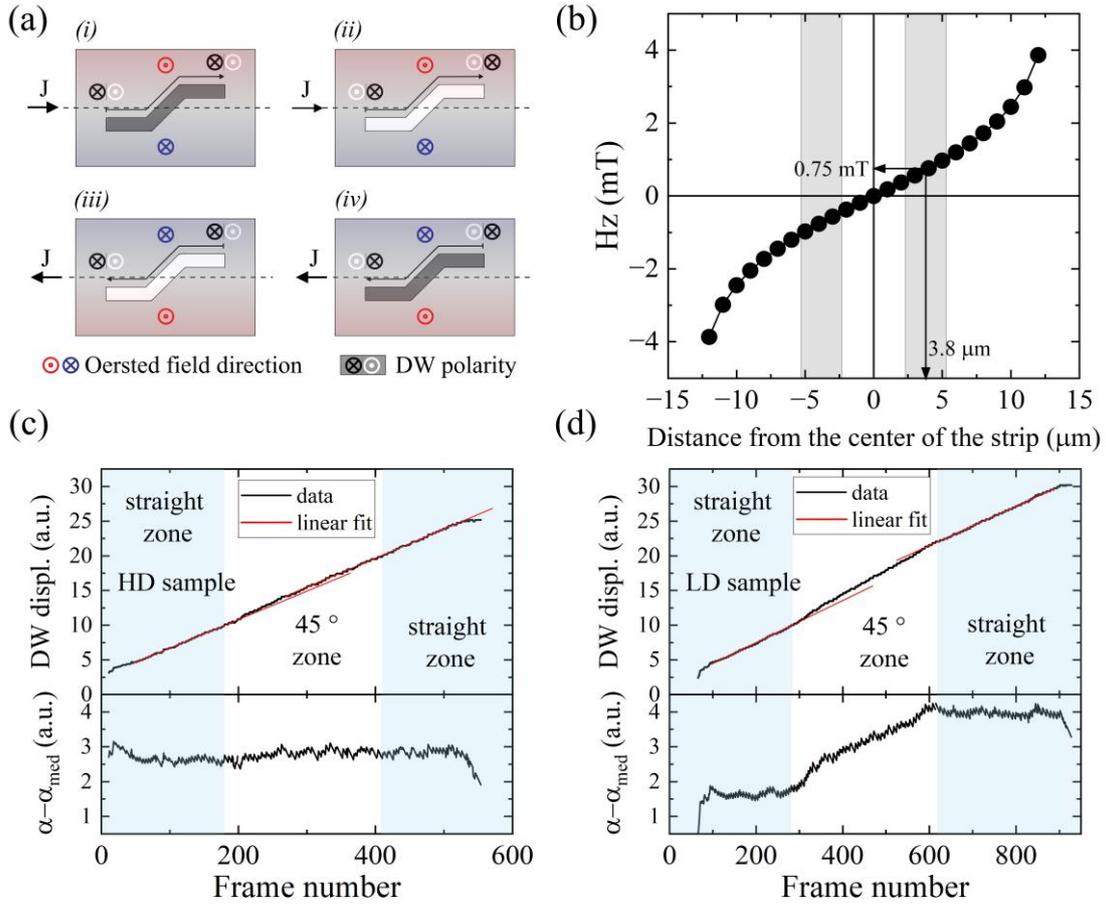

**Figure S11.** (a) Schematic representation of different polarity DWs motion under positive and negative current pulses. The out-of-plane Oersted field direction is also indicated. (b) The out-of-plane Oersted field as a function of distance from the center of the current strip, calculated at 0.3 nm above the strip for a current density of $10^{12}$ A/m². The gray-shaded areas in represent the position of the DW conduit straight zones. (c) and (d) the DW displacement versus the number of recorded frames for the HD and LD samples. The DW displacements in the two straight zones were linearly fitted and the average slope $[= (\alpha_{szI} + \alpha_{szII})/2]$ was extracted from the DW displacements values ($d_{DW} - \alpha_{med} \times$ frame number) and plotted in the lower part of the figures.

The fact that the DW velocities are the same in both straight zones suggests that the effect of the out-of-plane Oersted field is either negligible or below our detection limit. In Fig. S11(b), we present the out-of-plane Oersted field as a function of distance from the center of the current strip, calculated at 0.3 nm above the strip for a current density of $10^{12}$ A/m², using the relation provided in [15]. It should be noted that, for a given current density, the total current scales with the strip width, making the out-of-plane Oersted field independent of the strip width. The gray-shaded areas in Fig. S11(b) represent the position of the DW conduit straight zones. Although direct comparison is challenging, we estimate that the DL effective field driving the DW motion is at least one order of magnitude larger than the out-of-plane Oersted field [12], making its effect negligible. We believe this is the main reason we did not observe a velocity variation between the two straight zones. If the DW conduit were positioned closer to the edge of the Pt strip, the out-of-plane Oersted field would increase and could potentially influence DW velocity, though we expect this effect to remain minimal. Nevertheless, it would not impact the functionality of the device, as the out-of-plane Oersted field goes to zero in the middle of the 45° zone and would not affect the DW velocity in that region.

## S8. Different ferromagnetic material

An important question is whether the observed behavior is intrinsic or if it is an extrinsic material-dependent feature related to DW pinning. To investigate this, we deposited samples similar to those presented in the manuscript, but using Co instead of CoFeB. Co has a higher depinning field, and its crystalline structure results in more defects compared to the amorphous CoFeB. Consequently, Co exhibits a higher density and stronger pinning sites relative to CoFeB.

Figure S12 (a) and (b) show the motion of a down/up DW subjected to 3000 positive and negative pulses in a low DMI Si/SiO2/Ta(25)/Pt(60)/Co(6)/Pt(16) sample. When positive pulses are applied, the down/up DW moves through the entire structure, whereas with negative pulses, the DW motion is blocked upon entering the 45° zone. In comparison, for a high DMI Si/SiO2/Ta(25)/Pt(60)/Co(6)/Pd(4)/Pt(12) sample, as shown in Fig. S12 (c) and (d), the down/up DW moves through the entire structure after 2000 positive pulses. However, with the same number of negative pulses, the DW crosses the middle of the 45° zone.

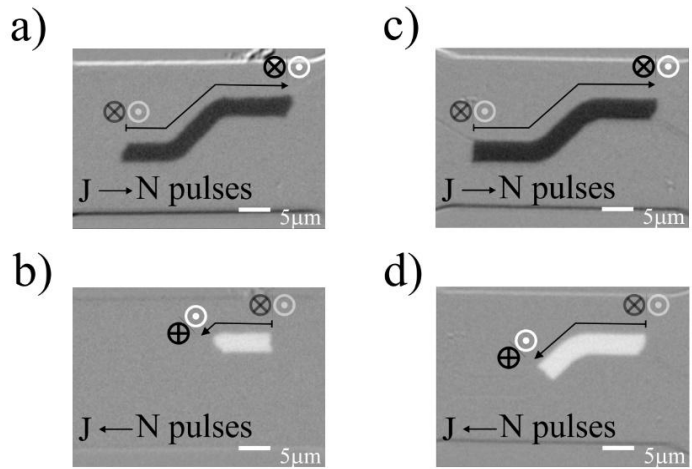

**Figure S12.** (a) and (b) the displacement of a down/up DW subjected to 3000 positive and negative pulses in a low DMI Si/SiO2/Ta(25)/Pt(60)/Co(6)/Pt(16) sample. (c) and (d) the displacement of a down/up DW subjected to 2000 positive and negative pulses in a high DMI Si/SiO2/Ta(25)/Pt(60)/Co(6)/Pd(4)/Pt(12) sample.

These results suggest that, similar to the CoFeB samples, the DW motion in the low DMI samples is blocked in the 45° zone, while for the high DMI samples, the DW velocity is reduced within this zone. This supports the idea that the observed behavior is intrinsic rather than being a material-dependent feature.